# PITCH CONTROL BY LQR FOR FIXED WING AIRCRAFT DURING MICROBURST ENCOUNTER

Şükrü Ayyıldız[1], Hakan Yazıcı[1,*]

**ABSTRACT**

In this study, a linear mathematical model representing longitudinal flight dynamics of an airplane is developed and responses of the aircraft during a microburst encounter are investigated. The effects of microburst that are acting on the aircraft are attempted to be suppressed with the elevator control surface of the aircraft, which is controlled by Linear Quadratic Regulator method. To illustrate the effectiveness of the proposed control method, numerous numbers of simulation studies is performed. As a result of the simulations, it is observed that effects of microburst on the pitching angle and altitude are significantly attenuated by the proposed control method. In addition, it is confirmed that the elevator control surface movement which provides the necessary controller input is within the physical capabilities of the aircraft.

**INTRODUCTION**

Modern airplanes and fixed wing unmanned aircraft use advanced flight control systems to reduce or even fully undertake the workload of pilots or other flight crew members by achieving tasks such as stability augmentation under severe atmospheric conditions. The early flight control systems developed using classical control methods have undertaken tasks such as speed, direction and altitude tracking. Although the conventional methods such as root locus and Bode plots, invented by Evans and Bode respectively, are very practical to many control applications for their simplicity and ease of use, the advantages of these methods are rapidly lost as the complexity of the systems increases. Optimal control methods aim to find an optimization criterion for engineering applications where classical control theory is insufficient and they have become easily solvable with developing computers.

As the air transportation became widespread around the world, the necessity of strict flight safety rules is emerged. Even today, severe atmospheric conditions are among the most important factors that is threatening flight safety. Airplanes encounter variable winds regularly during flight but since most of the flight goes within high cruising airspeeds and altitudes, these conditions don't cause strong deviations from intended flight paths, thus, an immediate danger. Variable winds that is changing their directions and severity with short periods of time possess danger to the airplanes at the most invulnerable phase of the flight, which is the landing. Under effects of atmospheric disturbances during landing, airplanes likely to have restricted opportunity to recover due to low altitude, airspeed and power. Among the different types of severe atmospheric conditions, microbursts are one of the most notorious dangers to the airplanes. A microburst can be defined as an intense downdraft wind that is collapsing on a small area and spreading around all directions once it hits to the ground, as shown in Figure 1.

An aircraft encounters a microburst first experiences a sudden increase of airspeed due to headwind on the outflow region, which causes an altitude gain. The pilot may try to reduce power and pitching angle to follow intended glideslope path, but headwind disappears and leaves its place to a tailwind after passing the core region of downdraft. The downdraft decreases the reference glideslope path angle and the tailwind decrease airspeed, thus lift. This causes a severe altitude loss and a high risk of a sudden crash to the ground. Even without an intervention from pilot, large amount of deviations from reference glideslope path is still may be experienced. Between the years 1964 and 1985, microbursts contributed in at least 26 civil aviation accidents involving almost 500 fatalities [1].

[1] Department of Mechanical Engineering, Yildiz Technical University, Istanbul, Turkey
*E-mail address: sukru.ayyildiz@std.yildiz.edu.tr, hyazici@yildiz.edu.tr
Orcid id: https://orcid.org/0000-0002-7716-7410, https://orcid.org/0000-0001-6859-9548

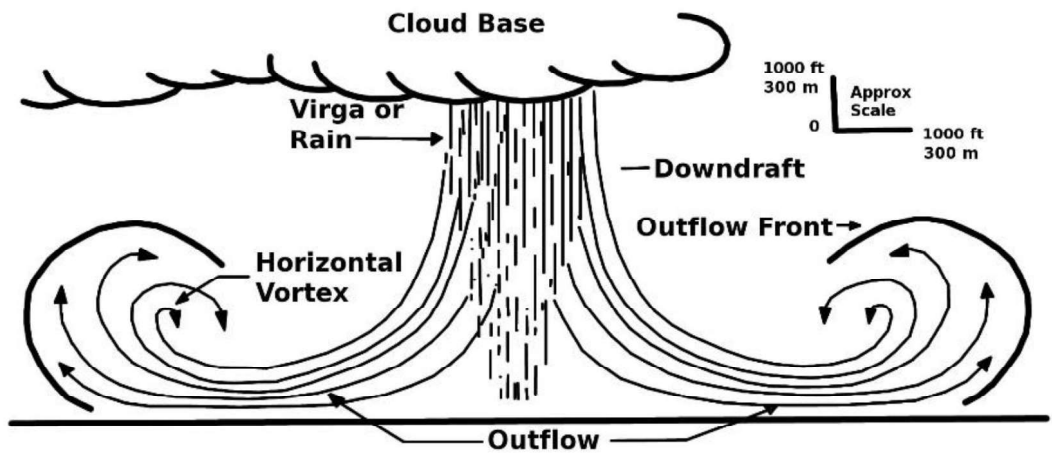

**Figure 1.** Wind profile for a symmetric microburst [2]. Reprinted from *Pilot Windshear Guide (p. 8)*, by D. C. Beaudette, 1988, Federal Aviation Administration Rept. AC-0045. In public domain.

Several research studies are conducted in recent decades to investigate the effects of a microburst on flight conditions [3, 4, 5]. Non-linear optimal control methods are proposed by Psiaki and Stengel [6, 7] to calculate optimum path for safely penetrating a microburst. The conclusion of their studies was a nonlinear, optimal flight path trajectory may provide both of jet transport and general aviation aircraft to penetrate microbursts with excellent desired flight path tracking. Although it's not contributing to the flight patch tracking as quickly as elevator input, in some cases throttle input proved itself as a helpful contributor to preventing a stall during transit phase in a microburst. Also, it is observed that optimally controlled jet transport aircraft can penetrate microbursts more successfully than lighter general aviation aircraft.

Dogan and Kabamba [8] developed microburst escape strategies by using sample analysis and statistical approach methods and they concluded that altitude and dive tracking may provide better results than pitching angle tracking. Pourtakdoust et al. [9] proposed a non-linear optimal trajectory planning method by coupling longitudinal and lateral – directional motions of the aircraft. Also, they indicate that they did not achieve a perfect close loop controller due to uncertainties about atmosphere, microburst and aircraft itself. All these studies have a common theme that providing the required correcting control input is likely to be a more dominant factor to successfully penetrate microbursts, rather than aircraft's physical performance capabilities.

In contrast to the many of the research studies in the literature, this study is based on a linearized mathematical model of longitudinal flight dynamics, which eliminates the difficulty of designing non-linear controllers. Also, this study proposes control via Linear Quadratic Regulator method, which is one of the fundamental problems of optimal control theory and frequently used due to its ease of development and ability to limit states and control inputs, unlike conventional control methods.

**THEORY**

In order to conduct simulation studies, it is required to assembly a proper mathematical model for the related system. In complex dynamical systems such as the aircraft discussed in this study, mathematical models become critical due to difficulties and costliness about prototyping and experimenting. In this study, the airplane is considered as a rigid body that has six degrees of freedom which describes three translational and three rotational motion. The longitudinal dynamics deal with rotational motion around $y_b$ axis (also called pitch axis), which is called "pitching" motion and translational motion along $x_b$ and $z_b$ axes (also called roll and yaw axis, respectively). The notation for the axes of body fixed coordinate system, translational velocities, angular velocities, forces, moments, system states, control inputs and atmospheric disturbance inputs can be seen in Figure 2.

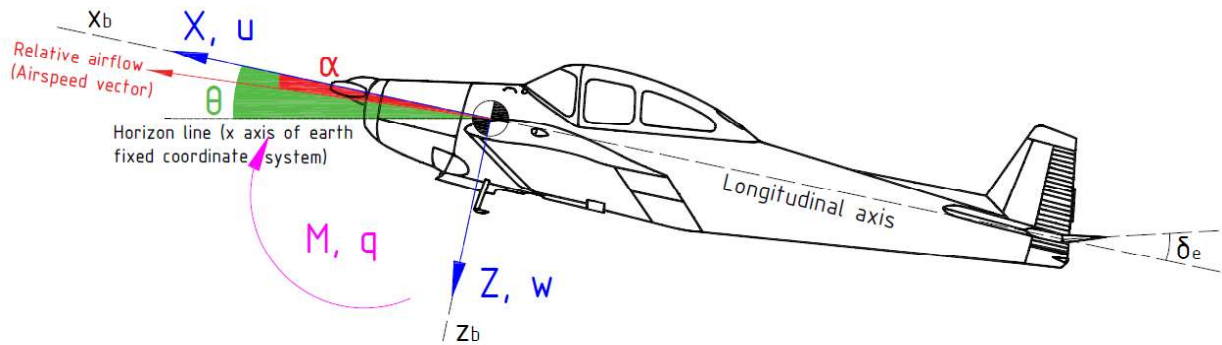

**Figure 2.** Visual representation of dynamical variables.

In Figure 2, $X$ and $Z$ are aerodynamic forces and $u$ and $w$ are airspeeds along body axes $x_b$ and $z_b$ respectively. $M$ is aerodynamic moment about $y_b$ axis. $\alpha$ is angle of attack, $\theta$ is pitching angle and $q$ is pitching rate. $\delta_e$ is elevator deflection.

In order to apply the proposed control method, the mathematical model must be linear. A highly accurate mathematical model for linearized longitudinal dynamics of an airplane can be evaluated by modifying the models included in previous studies and textbooks [10]. In addition, wind disturbance effects act on the aircraft in the same way as forward velocity and angle of attack do, so these are added to the dynamical equations in the same manner. In addition to 4 differential equations in the model, the 5$^{th}$ equation is also added to observe altitude, which can be expressed as:

$$\Delta \dot{h} = w = u_0 \sin(\Delta \alpha - \Delta \theta) \tag{1}$$

where $h$ is altitude. One can use $\sin\theta = \theta$ approximation to reduce Equation (1) to following:

$$\Delta \dot{h} = w = u_0 (\Delta \alpha - \Delta \theta) \tag{2}$$

This approximation has an error value less than %1 if the angles involved do not exceed 14°. This study investigates oscillatory deviations from steady cruising state, so it should be noted that all equations of motion in this mathematical model won't work properly for investigating aggressive flight maneuvers or extreme aerodynamic conditions such as flat spins and stalls. Final set of differential equations that is representing the longitudinal flight dynamics can described as below, in Equations (3-7):

$$\Delta \dot{u} = X_u \Delta u + X_\alpha \Delta a - g \Delta \theta + X_{\delta_e} \Delta \delta_e - X_u \Delta u_g - X_\alpha \Delta \alpha_g \tag{3}$$

$$\Delta \dot{\alpha} = \frac{Z_u}{u_0}\Delta u + \frac{Z_\alpha}{u_0}\Delta \alpha + \Delta q + X_{\delta_e}\Delta \delta_e - \frac{Z_u}{u_0}\Delta u_g - \frac{Z_\alpha}{u_0}\Delta \alpha_g \tag{4}$$

$$\Delta \dot{q} = M_u \Delta u + M_\alpha \Delta a + M_q \Delta_q + M_{d_e}\Delta \delta_e - M_u \Delta u_g - M_\alpha \Delta \alpha_g \tag{5}$$

$$\Delta \dot{\theta} = q \tag{6}$$

$$\Delta \dot{h} = u_0(\Delta \alpha - \Delta \theta) \tag{7}$$

Mathematical model can be expressed in state – space representation in the following manner:

$$\dot{x} = Ax + Bu + G\eta \qquad (8)$$
$$y = Cx + Du \qquad (9)$$

where $x$ is state vector, $u$ is control input vector, $\eta$ is disturbance input vector, $y$ is output vector, $A$ is state matrix, $B$ is control input matrix, is $G$ disturbance input matrix, $C$ is output matrix and $D$ is feedforward matrix. Equations (3-7) can be written in state – space matrix form as follows:

$$\begin{bmatrix} \Delta\dot{u} \\ \Delta\dot{\alpha} \\ \Delta\dot{q} \\ \Delta\dot{\theta} \\ \Delta\dot{h} \end{bmatrix} = \begin{bmatrix} X_u & X_\alpha & 0 & -g & 0 \\ \frac{Z_u}{u_0} & \frac{Z_\alpha}{u_0} & 1 & 0 & 0 \\ M_u & M_\alpha & M_q & 0 & 0 \\ 0 & 0 & 1 & 0 & 0 \\ 0 & u_0 & 0 & -u_0 & 0 \end{bmatrix} \begin{bmatrix} \Delta u \\ \Delta \alpha \\ \Delta q \\ \Delta \theta \\ \Delta h \end{bmatrix} + \begin{bmatrix} X_{\delta_e} \\ Z_{\delta_e} \\ M_{\delta_e} \\ 0 \\ 0 \end{bmatrix} [\Delta\delta_e] + \begin{bmatrix} -X_u & -X_\alpha \\ -\frac{Z_u}{u_0} & -\frac{Z_\alpha}{u_0} \\ -M_u & -M_\alpha \\ 0 & 0 \\ 0 & 0 \end{bmatrix} \begin{bmatrix} \Delta u_g \\ \Delta \alpha_g \end{bmatrix} \qquad (10)$$

$$y = I_{5x5} \begin{bmatrix} \Delta u & \Delta\alpha & \Delta q & \Delta\theta & \Delta h \end{bmatrix}^T + [0]u \qquad (11)$$

These equations include *stability derivatives*. The method of using stability derivatives are a very reliable approach for linearizing flight dynamics [11] and they express how particular forces and moments change around a steady (*trim*) condition with respect to different flight parameters. For instance, $Z_\alpha$ parameter in Equation (4) describes how the Z force changes with respect to angle of attack $\alpha$, around the trim condition. Also, parameters with index $\delta_e$ are *control derivatives* and evaluated in the same manner – only difference is they express the change in the main parameter with respect to elevator control input, not respect to a flight parameter as stability derivatives.

To conduct simulation studies, all the parameters in Equation (10) must have their numerical values. Stability derivatives are evaluated according to aerodynamic data of the aircraft. Obtaining the aerodynamic data is beyond the scope of this study because it requires to perform live flight tests, wind tunnel tests and computational fluid dynamics simulations [12]. Instead, they are taken directly from results of these tests and simulations [13] and stability derivatives are calculated according to these aerodynamic data. Single – engine, four – seat, light aircraft [14] *Ryan Navion* is selected for this study due to its ease to obtaining aerodynamic data, which is shared with the aeronautics community for a long time. All the calculated stability derivatives with their values are can be seen in Table 1.

Table 1. Calculated stability derivatives and their numerical values.

| Stability derivative | Value | Unit |
|---|---|---|
| $X_u$ | -0.0454 | $s^{-1}$ |
| $X_\alpha$ | 1.9609 | $m/s^2$ |
| $X_q$ | 0 | $s^{-1}$ |
| $X_{\delta_e}$ | 0 | $m/s^2$ |
| $Z_u$ | -0.3722 | $s^{-1}$ |
| $Z_\alpha$ | -116.9207 | $m/s^2$ |
| $Z_q$ | 0 | $s^{-1}$ |
| $Z_{\delta_e}$ | -8.7016 | $m/s^2$ |
| $M_u$ | 0 | $1/ms$ |
| $M_\alpha$ | -8.9246 | $s^{-2}$ |
| $M_q$ | -2.0968 | $s^{-1}$ |
| $M_{\delta_e}$ | -12.0606 | $s^{-2}$ |

If the numerical values given in Table 1 are plugged into Equation (10), state − space matrix form of mathematical model with numerical values can be obtained as follows:

$$\begin{bmatrix}\Delta\dot{u}\\ \Delta\dot{\alpha}\\ \Delta\dot{q}\\ \Delta\dot{\theta}\\ \Delta\dot{h}\end{bmatrix}=\begin{bmatrix}-0.0454 & 1.9609 & 0 & -9.8066 & 0\\ -0.0069 & -2.1652 & 1 & 0 & 0\\ 0 & -8.9246 & -2.0968 & 0 & 0\\ 0 & 0 & 1 & 0 & 0\\ 0 & 54 & 0 & -54 & 0\end{bmatrix}\begin{bmatrix}\Delta u\\ \Delta\alpha\\ \Delta q\\ \Delta\theta\\ \Delta h\end{bmatrix}+\begin{bmatrix}0\\ -0.1611\\ -12.0606\\ 0\\ 0\end{bmatrix}[\Delta\delta_e]+\begin{bmatrix}0.0454 & -1.9609\\ 0.0069 & 2.1652\\ 0 & 8.9246\\ 0 & 0\\ 0 & 0\end{bmatrix}\begin{bmatrix}\Delta u_g\\ \Delta\alpha_g\end{bmatrix} \quad (12)$$

Next step is designing the controller. Linear quadratic regulator (LQR) method is one of the main problems of control theory that aims to operate a dynamic system in the most effective way at minimum cost. It requires that the system dynamics to be described by a set of linear equations and the cost by a quadratic function. Before obtaining control law, one must check the system's controllability and observability. Controllability is a property of linear systems that concerns with whether the control input can affect all states of the system. System in this study is of $5^{th}$ order, so if the controllability matrix $C_M$, defined as:

$$C_M = [\ B\quad AB\quad A^2B\quad A^3B\quad A^4B] \quad (13)$$

is of rank 5, it can be said that this system is controllable. On the other hand, observability is a property of linear systems that concerns with whether the states of the input can be identified from output of the system. If the observability matrix $O_M$, defined as:

$$O_M = [\ C\quad CA\quad CA^2\quad CA^3\quad CA^4]^T \quad (14)$$

is of rank 5, it can be said that this system is observable. Due to high order of the system, it is difficult to calculate the controllability and observability matrices by hand. MATLAB software is used to conduct this task and rank of the both matrices are obtained as 5, which indicates the system is controllable and observable.

The cost function is a quadratic function which drives the system states from initial time $t_0$ to final time $t_f$, with respect to desired performance and cost criteria. The quadratic cost function can be defined by Equation (15):

$$J = \int_{t_0}^{t_f}[x^T(t)Qx(t) + u^T(t)Ru(t)]dt \quad (15)$$

where $Q$ is the weighting matrix of the states, $R$ is the weighting matrix of the control input and $J$ is the cost function. The $Q$ and $R$ matrices are diagonal square matrices that specify the desired control amount of the states and controller penalty, respectively. For many real-world applications, it may be required to include a penalty for controller input due to physical constraints such as angular motion limit of the actuator (which is the case in this study) or spent energy.

To develop a linear quadratic regulator, a controller gain must be found to minimize the cost while ensuring the control input of the system is feasible or physically possible to apply. The optimal control law as the feedback of all states in the linear system can be described as follows:

$$u = -Kx \quad (16)$$

where $K$ is the unknown gain matrix and $x$ is the state vector. To design a linear quadratic regulator, one must calculate the $K$ matrix. If the cost function in Equation (15) is solved, a Riccati differential equation can be obtained:

$$\frac{dS(t)}{dt} = S(t)BR^{-1}B^TS(t) - S(t)A - A^TS(t) - Q \quad (17)$$

where $S(t)$ is the symmetric, positive definite Riccati matrix. Time varying gain matrix $K(t)$ can be obtained by solving the equation below:

$$K(t) = R^{-1}B^T S(t) \tag{18}$$

If the final time $t_f$ approaches the infinity, time varying Riccati matrix becomes a constant matrix and one can obtain reduced form of Equation (17) as a nonlinear algebraic equation, which can be expressed as below:

$$SBR^{-1}B^T S - SA - A^T S - Q = 0 \tag{19}$$

Unless for simple cases, solution of this Riccati equation requires computer software. Weighting matrices $Q$ and $R$ are determined as below:

$$Q = \begin{bmatrix} 0 & 0 & 0 & 0 & 0 \\ 0 & 150 & 0 & 0 & 0 \\ 0 & 0 & 0 & 0 & 0 \\ 0 & 0 & 0 & 2000 & 0 \\ 0 & 0 & 0 & 0 & 0.01 \end{bmatrix} \tag{20}$$

$$R = [30] \tag{21}$$

MATLAB software is used to calculate controller gain matrix, as it is very difficult to solve Equation (19) by hand for a 5$^{th}$ order system. Riccati matrix $S(t)$ and controller gain $K(t)$ matrices are obtained as below:

$$S = \begin{bmatrix} 0.6115 & -5.0662 & 0.1220 & 5.9464 & -0.2053 \\ -5.0662 & 117.1642 & -3.7795 & -132.8100 & 3.0884 \\ 0.1220 & -3.7795 & 2.4973 & 23.5294 & -0.0867 \\ 5.9464 & -132.8100 & 23.5294 & 436.7515 & -3.8089 \\ -0.2053 & 3.0884 & -0.0867 & -3.8089 & 0.1260 \end{bmatrix} \tag{22}$$

$$K = [-0.0219 \quad 0.8901 \quad -0.9837 \quad -8.7459 \quad 0.0183] \tag{23}$$

The magnitude and direction pattern of disturbance winds can be realized from Figure 1. Magnitude of gust wind along $x$ axis, $u_g$ increases as the aircraft enters the microburst, then decreases until reaching zero at the core of microburst. Then it will start to increase its magnitude with the reverse direction, followed by deteriorating until aircraft finally exits the microburst. On the other hand, magnitude of gust wind along $z$ axis, $w_g$ will going to increase from since the beginning to its peak, the core. Then it will start to deteriorate until the aircraft exits microburst completely. Two disturbance winds are graphically described in Figure 3. Also, these two disturbance effects are determined as sine functions, which stated below:

$$u_g = 3\sin(0.05t), \quad 0 \leq t \leq 20 \ s \tag{24}$$

$$w_g = -5\sin(0.025t), \quad 0 \leq t \leq 20 \ s \tag{25}$$

Magnitudes are in units of m/s, and frequencies are in units of Hz. It should be noted that one must simply divide $w_g$ to cruising speed $u_0$ to obtain $\alpha_g$ (disturbance angle of attack, $\alpha_g$ caused by $w_g$) to plug it in Equation (10), since the state - space form of mathematical models requires $\alpha_g$ instead of $w_g$.

$$\alpha_g = \frac{w_g}{u_0} = \frac{w_g}{54} \tag{26}$$

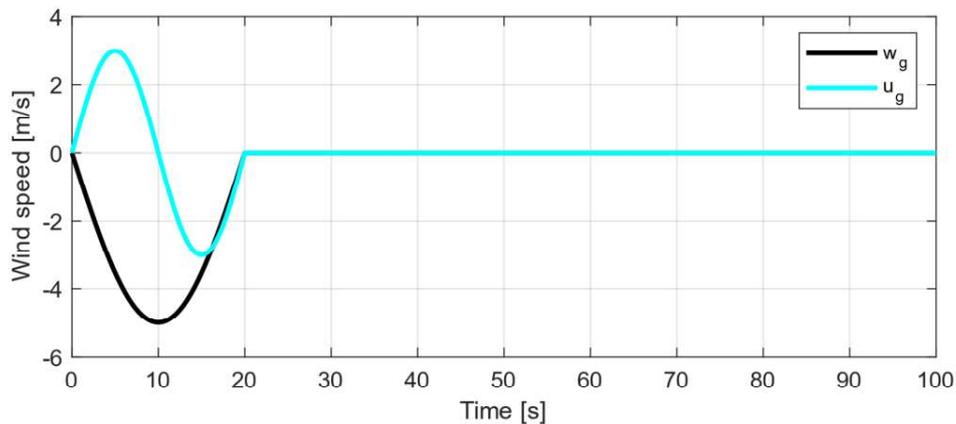

**Figure 3.** Disturbance wind inputs $u_g$ and $w_g$ over time.

Final step is conducting the simulation studies by using MATLAB and Simulink software to observe uncontrolled and controlled system responses under disturbance effects and the controller input.

**RESULTS AND DISCUSSION**

Before evaluating system responses, it will be useful to see the two distinct natural frequencies of two longitudinal modes. The natural frequency of short period mode is 0.584 Hz and phugoid (long period) mode is 0.033 Hz. Figure 4 shows the pitching angle response of the aircraft at uncontrolled condition and controlled by LQR.

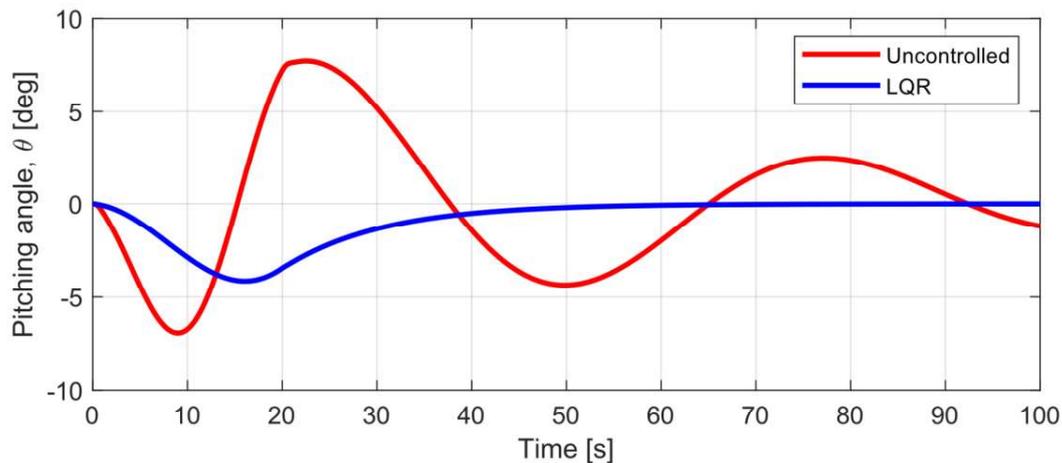

**Figure 4.** Pitching angle $\theta$ over time.

It can be clearly seen that given $w_g$ disturbance excited the phugoid mode of the aircraft to some extent. Uncontrolled pitching angle response enters phugoid motion, which is one of the basic aircraft motions that indicates a slow exchange between kinetic and potential energies. In phugoid motion, aircraft pitches down, gains airspeed and lift force increases. Due to increased lift force, aircraft begins to pitch up and lose airspeed, thus lift force. Due to decreased lift force, aircraft pitches down again and this cycle goes on until the oscillation completely dampens out. It must be noted that this motion is relatively easy to be corrected by pilots but due to its long period, but it can cause very troublesome events during landing. At the uncontrolled response, although it's stable, phugoid motion did not dampen out completely even until 100 seconds of simulation time. The maximum values of pitching angle are can be observed as 7.7 degrees as maximum and −7 degrees as minimum.

On the other hand, if the LQR − controlled pitching angle response is investigated, the maximum deviation from steady condition is observed as −4.1 degrees. Also, the response does not pass above the steady condition of 0 degrees and the oscillation is eliminated. Lastly, the system response settles on steady state at near 30 seconds mark.

Quite severe effects are observed on the uncontrolled altitude response, which can be seen in Figure 5. It is observed that aircraft loses a maximum altitude of 108.9 meters, then settles itself around 60 meters below of cruising altitude. It must be noted that aircraft's altitude never returns to its trim condition 0 meters. The reason behind this occurrence is simple: just after the 10 seconds mark, which is just after the aircraft's pass beyond the microburst's core region, the exact occurrence that mentioned previously in happened in introduction section. After passing the core of the microburst, aircraft lost its altitude drastically. This observation reveals another hazardous condition: if the aircraft's cruising altitude is below of 108.9 meters, aircraft will crash into the ground.

If the LQR − controlled altitude response is investigated, the maximum altitude loss is observed as 26.3 meters. Although it's not zero, this result is way safer that uncontrolled response, which satisfactorily increases the safety margin of this microburst transit.

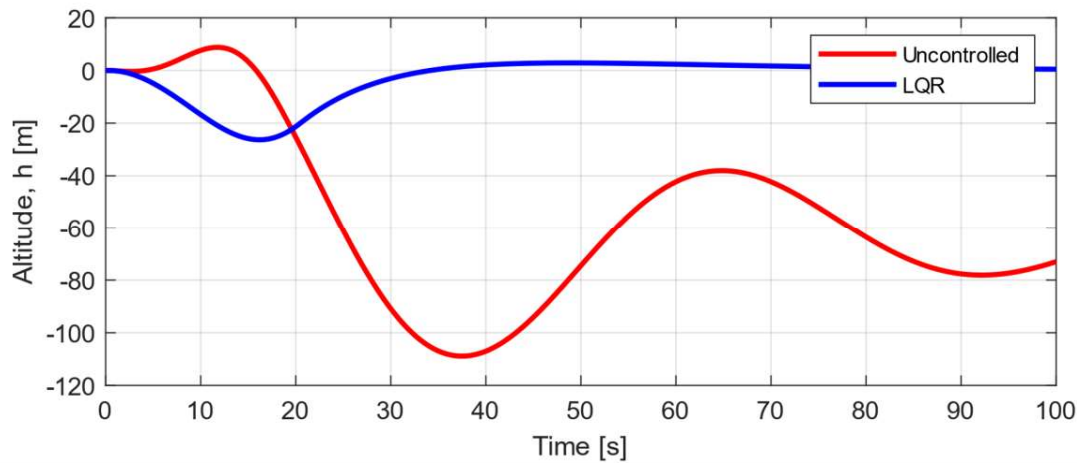

**Figure 5.** Altitude $h$ over time.

The last observation is about elevator control surface input to achieve simulated controlled responses. It is observed that the elevator control surface deflection angle is between −0.47 and 0.85 degrees. The results are well within the physical deflection capability of the elevator control surface in the aircraft, which is usually between −20 and 20 degrees. The plot of elevator deflection over time can be seen in Figure 6.

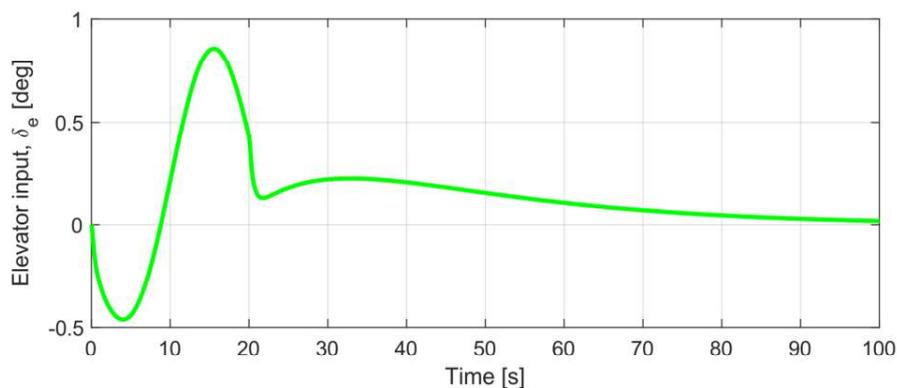

**Figure 6.** Elevator control surface input $\delta_e$ over time.

## CONCLUSION

In order to attenuate the effects of a microburst transit on a general aviation light aircraft, a linear quadratic regulator is developed by determining weighting matrices according to desired performance results. The system states are focused on this study are pitching angle and aircraft altitude. Pitching angle is a comfort condition as its effect can be felt directly by people onboard on the aircraft, but also it contributes the flight safety as excessive values of pitching angle can cause unexpected and hazardous consequences. On the other hand, altitude is purely a safety factor which has very low tolerance to unwanted behavior. First, uncontrolled behavior of pitching angle and altitude states are evaluated. It is observed from the pitching angle that aircraft's phugoid dynamic mode is excited by disturbance wind inputs and aircraft enters the phugoid motion, begins to oscillate with a period of approximately 30 seconds. On the other hand, maximum altitude loss is observed as $108.9$ meters, which is dangerous for a general aviation aircraft especially if its altitude is low.

Proposed controller improved both responses significantly. The pitching angle oscillation is eliminated and maximum pitching angle deviation is reduced by $\%46.8$. Maximum altitude loss is decreased to $26.3$ from $108.9$ meters, which means it's reduced by $\%75.8$. In addition, required elevator control surface input is observed as between $-0.47$ and $0.85$ degrees, which is well within the physical deflection capability of the elevator control surface (usually between $-20$ and $20$ degrees) of the aircraft.

## NOMENCLATURE

| | |
|---|---|
| $x_b, y_b, z_b$ | Axes of body fixed coordinate system. |
| $u$ | Forward speed, m/s. |
| $w$ | Vertical speed, m/s. |
| $q$ | Pitching rate, deg/s. |
| $h$ | Altitude, m. |
| $g$ | Gravitational acceleration, m/s$^2$. |
| $X$ | Aerodynamic force along the $x_b$ axis, N. |
| $Z$ | Aerodynamic force along the $z_b$ axis, N. |
| $M$ | Aerodynamic moment along the $y_b$ axis, Nm. |
| $X_\bullet$ | Stability derivative of aerodynamic force along the $x_b$ axis with respect to ●. |
| $Z_\bullet$ | Stability derivative of aerodynamic force along the $z_b$ axis with respect to ●. |
| $M_\bullet$ | Stability derivative of aerodynamic moment around $y_b$ axis with respect to ●. |
| $x$ | State vector |
| $u$ | Control input vector |
| $y$ | Output vector |
| $\eta$ | Disturbance input vector |
| $J$ | Cost function |
| $Q$ | Weighting matrix of the states |
| $R$ | Weighting matrix of the control input |
| $C_M$ | Controllability matrix |
| $O_M$ | Observability matrix |

Greek symbols

| | |
|---|---|
| α | Angle of attack, deg. |
| $\theta$ | Pitching angle, deg. |
| $\delta$ | Control surface deflection, deg. |
| $\Delta$ | Deviation from steady state. |
| $\eta$ | Disturbance input vector |

Subscripts

| | |
|---|---|
| $g$ | Refers to gust (i.e. disturbance wind) |
| $e$ | Refers to elevator control surface |
| 0 | Refers to trim condition (i.e. cruise, steady state) and refers to initial |
| $f$ | Refers to final |

**REFERENCES**

[1] National Research Council. (1983). *Low-altitude wind shear and its hazard to aviation*. National Academies Press.
[2] Beaudette, D. C. (1988). Pilot windshear guide (p. 8). *Federal Aviation Administration Rept. AC00-45*.
[3] Frost, W., & Bowles, R. L. (1984). Wind shear terms in the equations of aircraft motion. *Journal of Aircraft*, *21*(11), 866-872.
[4] Miele, A., Wang, T., & Melvin, W. (1987). Optimization and acceleration guidance of flight trajectories in a windshear. *Journal of Guidance, Control, and Dynamics*, *10*(4), 368-377.
[5] Mulgund, S. S., & Stengel, R. F. (1993). Target pitch angle for the microburst escape maneuver. *Journal of Aircraft*, *30*(6), 826-832.
[6] Psiaki, M. L., & Stengel, R. F. (1991). Optimal aircraft performance during microburst encounter. *Journal of guidance, control, and dynamics*, *14*(2), 440-446.
[7] Psiaki, M. L., & Stengel, R. F. (1986). Optimal flight paths through microburst wind profiles. *Journal of Aircraft*, *23*(8), 629-635.
[8] Dogan, A., & Kabamba, P. T. (2000). Escaping microburst with turbulence: altitude, dive, and pitch guidance strategies. *Journal of Aircraft*, *37*(3), 417-426.
[9] Pourtakdoust, S. H., Kiani, M., & Hassanpour, A. (2011). Optimal trajectory planning for flight through microburst wind shears. *Aerospace Science and Technology*, *15*(7), 567-576.
[10] Nelson, R. C. (1998). *Flight stability and automatic control* (Vol. 2, p. 105). New York: WCB/McGraw Hill.
[11] Bryan, G. H. (1911). *Stability in aviation: an introduction to dynamical stability as applied to the motions of aeroplanes*. Macmillan and Company, limited.
[12] Williams, J. E., & Vukelich, S. R. (1979). *The usaf stability and control digital datcom. volume ii. implementation of datcom methods*. MCDONNELL DOUGLAS ASTRONAUTICS CO ST LOUIS MO.
[13] Schmidt, D. K. (2012, pp. 822 – 823). *Modern flight dynamics*. New York: McGraw-Hill.
[14] Simpson, R. (2005). *The general aviation handbook*. Midland Publishing.